\documentclass[twocolumn,nofootinbib,preprintnumbers,floatfix,aps,prl,10pt]{revtex4-2}

\usepackage[T1]{fontenc}
\usepackage{amsmath,amssymb,bm}
\usepackage{booktabs}
\usepackage{graphicx}
\usepackage{placeins} %
\usepackage{hyperref}
\hypersetup{hidelinks}

\newcommand{\as}{\alpha_s}
\newcommand{\Nc}{N_c}
\newcommand{\Tr}{\operatorname{Tr}}
\newcommand{\Gc}{\Gamma^c}
\newcommand{\Gb}{\bar\Gamma}

\begin{document}

\title{A First Glimpse at Super-Leading Logarithms in Monte Carlo Evolution}

\author{Nicolas Schalch}
\affiliation{Rudolf Peierls Centre for Theoretical Physics, Clarendon Laboratory,
Parks Road, University of Oxford, Oxford OX1 3PU, United Kingdom}

\begin{abstract}
Super-leading logarithms from Glauber exchange complicate the
resummation of non-global observables at hadron colliders. In this Letter, we extend
Marzili, a Monte Carlo framework for non-global resummation based on
effective field theory, to resum the leading nonzero colour contribution
involving two Glauber exchanges.
This is the first Monte Carlo resummation of super-leading logarithms.
Together with Marzili's treatment of non-global logarithms, this opens a path
towards studying their interplay and incorporating Glauber exchange
into parton showers.

\end{abstract}

\maketitle

\emph{Introduction.---}
Measurements that restrict radiation at a scale $Q_0$ far below
the hard scale $Q$ probe correlations across widely separated scales.
Such restrictions
prevent complete cancellation between real and virtual corrections.
Integrations over soft and collinear radiation then leave large logarithms
$L=\ln(Q/Q_0)$ in the perturbative coefficients. Powers of these logarithms
can compensate the smallness of $\as$, so a reliable prediction requires
summing the enhanced terms to all orders.

A canonical example is a rapidity gap: radiation above the veto
scale $Q_0$ is excluded from a chosen rapidity interval, while radiation
outside it is allowed. Partons outside the gap can radiate softer gluons
into it. These correlations across the gap boundary generate non-global
logarithms (NGLs)~\cite{Dasgupta:2001sh,Dasgupta:2002bw,Banfi:2002hw}.
Each emission can become a new radiator, so resummation requires
control over correlations among arbitrarily many partons. Extending
this resummation beyond leading-logarithmic (LL) accuracy was a
longstanding challenge, with next-to-leading logarithmic (NLL)
predictions becoming available only recently~\cite{Banfi:2021owj,Banfi:2021xzn,Becher:2021urs,Becher:2023vrh,FerrarioRavasio:2023kyg,Becher:2026zon}.
These NLL resummations rely on the large-$\Nc$ approximation.
The Marzili framework~\cite{Becher:2023vrh}, which we extend in this work,
achieves this accuracy by
solving renormalization-group~(RG) equations derived in effective field
theory (EFT) with Monte Carlo methods.

At finite $\Nc$, quantum interference requires a richer colour description.
Approaches include Langevin evolution of Wilson
lines~\cite{Weigert:2003mm,Hatta:2013iba,Hatta:2017fwr,Hatta:2020wre},
amplitude evolution in a colour-flow basis and its implementation in
CVolver~\cite{Platzer:2013fha,AngelesMartinez:2018cfz,DeAngelis:2020rvq,
Platzer:2022jny,Forshaw:2025fif,Forshaw:2025bmo}, and systematic colour
corrections and virtual phases in Deductor~\cite{Nagy:2019rwb}.
Efficient colour prescriptions recover full-colour NLL accuracy
for global observables~\cite{Hamilton:2020rcu}. For NGLs they remain
approximate, yet give remarkably good agreement with the finite-colour
rapidity-slice calculation of~\cite{Hatta:2013iba}, within its
statistical precision.

In $e^+e^-$ collisions, non-global soft radiation generates a
single-logarithmic series, specifically $\as^n L^n$. At hadron colliders, Glauber phases spoil the
cancellation of initial-state collinear radiation and introduce a
double-logarithmic enhancement. The resulting super-leading logarithms
(SLLs) start at four loops, $\as^4L^5$~\cite{Forshaw:2006fk}, and form the
tower $\as^3L^3(\as L^2)^n$ with $n\geq1$~\cite{Becher:2021zkk,Becher:2023mtx}.
Their resummation requires colour interference and a consistent
ordering of emissions and Glauber exchanges. In a parton shower, this
ordering is set by an evolution variable, such as transverse momentum,
whose choice matters for reproducing the SLL series~\cite{AngelesMartinez:2018cfz}.
Their consistent incorporation into parton showers has therefore remained elusive.
An all-order resummation of the intertwined collinear and Glauber
evolution responsible for this SLL series was achieved only
recently using soft-collinear effective
field theory~\cite{Bauer:2000ew,Bauer:2000yr,Bauer:2001ct,
Bauer:2001yt,Beneke:2002ph,Beneke:2002ni}, first for quark-initiated
processes~\cite{Becher:2021zkk} and then for arbitrary incoming
partons~\cite{Becher:2023mtx}.
Subsequent work included higher
Glauber exchanges for quark- and gluon-initiated
processes~\cite{Boer:2023quark,Boer:2023gluon} and running-coupling
effects through RG-improved evolution~\cite{Boer:2024rg}.
The series with arbitrarily many Glauber exchanges can be resummed
in closed form for all incoming channels at the leading nonzero order
in the colour expansion~\cite{Boer:2024xzy}.

In this Letter, we extend Marzili's numerical colour evolution to resum
the SLL series involving two Glauber exchanges, with
arbitrarily many collinear emissions
and one final soft emission.
We numerically solve the RG equations governed by the anomalous
dimension of \cite{Becher:2021zkk,Becher:2023mtx} by generating
successive (collinear) emissions and local colour reconnections.
The colour representation builds
on colour-flow methods for amplitude
evolution~\cite{DeAngelis:2020rvq,Forshaw:2025bmo}.
Only a bounded number of reconnections between the
amplitude and conjugate-amplitude flows is needed, even as the emission
multiplicity grows. At fixed $\Nc\as$, we retain coefficients
scaling as $\as^{n+3}\Nc^{n+1}=(\Nc\as)^{n+3}/\Nc^2$.

For forward $qq'\to qq'$ and $qg\to qV$, with colourless $V$,
we validate our approach against
all-order analytic resummations.

\emph{EFT framework.---}
We consider hadron collisions producing hard jets and possibly
colourless particles, with a veto at a perturbative scale $Q_0\ll Q$ in
an angular region separated from the beams and hard jets \cite{Sterman:1977wj}. 
Radiation
outside this region is unrestricted.
To separate the hard scattering from radiation constrained by the
veto, we use an EFT for non-global
observables~\cite{Becher:2015hka,Becher:2016mmh}.
At leading power in $Q_0/Q$, the hadronic factorization formula is~\cite{Becher:2021zkk}
\begin{equation}
 \sigma(Q_0)=\sum_{a,b}\int dx_1dx_2\sum_{m\geq m_0}
 \big\langle\mathcal H_m\otimes\mathcal W_m\big\rangle .
 \label{eq:factorization}
\end{equation}
Here $a,b$ label the incoming parton species, $x_{1,2}$ their momentum
fractions, and $m_0$ the number of coloured Born legs, including the two
incoming partons. The hard functions $\mathcal H_m(\{n\},Q,\mu)$ are colour density matrices
for two incoming and $m-2$ outgoing coloured partons \cite{Becher:2015hka,Becher:2016mmh,Becher:2021zkk}, with energies integrated
at fixed directions; $\mathcal W_m$ contains Wilson lines, incoming
collinear fields and the veto measurement. The brackets denote a colour
trace and $\otimes$ angular integrations. At the accuracy considered,
$\mathcal W_m=f_a(x_1,\mu_s)f_b(x_2,\mu_s)\mathbf1$ at $\mu_s\sim Q_0$,
where $f_a$ are parton distribution functions (PDFs).
Perturbative Glauber contributions to the low-energy matrix elements
restore compatibility with PDF factorization below $Q_0$~\cite{Becher:2024kmk,Becher:2025igg}.
Although these low-energy matching corrections can contain rapidity
logarithms, they contribute below the leading powers
$\as^{n+3}L^{2n+3}$ retained here. The leading-order boundary condition
therefore suffices. We study the resulting partonic
coefficient normalized to $\hat\sigma_B=\langle\mathcal H_{m_0}\rangle$.

Writing $\ell=\ln(Q/\mu)$ and $a_s=\as/(4\pi)$, the one-loop
anomalous dimension governing the evolution towards lower scales is
\begin{equation}
 \mathcal K_\ell=a_s\bigl(\Gb+V_G-2\ell\Gc\bigr),
 \label{eq:generator}
\end{equation}
with the purely collinear PDF evolution factored out. In the convention of Ref.~\cite{Becher:2021zkk},
using incoming labels $0,1$,
\begin{align}
 \Gc&=4\sum_{i=0,1}\bigl[C_i\mathbf1
 -\bm T_{i,L}\circ\bm T_{i,R}\,\delta(n_k-n_i)\bigr],
 \label{eq:gc}\\
 V_G&=-8i\pi\bigl(\bm T_{0,L}\cdot\bm T_{1,L}
                  -\bm T_{0,R}\cdot\bm T_{1,R}\bigr).
 \label{eq:vg}
\end{align}
Here $C_i$ is the quadratic Casimir; $L,R$ act on amplitude and conjugate
amplitude. The real operator $\circ$ adds a gluon with uncontracted colour
indices, exactly along beam $i$. The angular delta function is normalized to
$d\Omega_k/(4\pi)$. The remaining operator $\Gb$ contains wide-angle real
and virtual radiation, with the collinear terms subtracted.

Collinear coherence and
$\langle\mathcal H\Gc\rangle=\langle\mathcal H V_G\rangle=0$ imply the
common colour structure~\cite{Becher:2021zkk,Becher:2023mtx}
\begin{equation}
 C_{rn}=\big\langle\mathcal H_{m_0}(\Gc)^rV_G
                   (\Gc)^{n-r}V_G\Gb\big\rangle,\quad 0\leq r\leq n,
 \label{eq:words}
\end{equation}
with $m_0=4$ for $qq'\to qq'$ and $m_0=3$ for $qg\to qV$.
Each collinear insertion supplies two logarithms, while a Glauber or soft
insertion supplies one. Thus the series is $\as^{n+3}L^{2n+3}$; $n$ counts
real and virtual collinear operators, not emitted particles. The
$n=0$ term is a three-loop phase contribution; the pure SLL enhancement starts
at $n=1$.

At fixed coupling, we introduce
\begin{equation}
 t=\frac{\sqrt{\as}}{4\pi}\ell,\hspace{5mm}
 T=\frac{\sqrt{\as}}{4\pi}L,\hspace{5mm} \xi=16\pi^2T^2=\as L^2 .
 \label{eq:variables}
\end{equation}
The one-loop anomalous dimension in terms of these variables is
$\mathcal K_t=\sqrt{\as}(\Gb+V_G)-8\pi t\Gc$, since 
$d\ell/dt=4\pi/\sqrt{\as}$.
Collecting the hard functions into a row vector $\mathcal H(t)$,
the RG equation is $d\mathcal H(t)/dt=\mathcal H(t)\mathcal K_t$.
Its solution $\mathcal H(t_b)=\mathcal H(t_a)U(t_a,t_b)$ involves
the path-ordered exponential
\begin{equation}
 U(t_a,t_b)=\mathcal P\exp\!\left[
                    \int_{t_a}^{t_b}ds\,\mathcal K_s\right].
 \label{eq:rg-solution}
\end{equation}
The ordering places earlier operators to the left when acting on
$\mathcal H$. This solution includes arbitrary collinear, soft and
Glauber insertions. To isolate the SLL series in \eqref{eq:words},
we retain exactly two powers of $V_G$ and one of $\Gb$, while resumming
$\Gc$ to all orders. We denote by $U_c$ the evolution generated by
$-8\pi t\Gc$ alone. Between the explicit insertions, only $U_c$ is needed.

Collinear coherence, $\mathcal H\Gc\Gb=\mathcal H\Gb\Gc$,
removes collinear evolution after the second Glauber from the final
trace. The cumulative coefficient
$C(T)\equiv\Delta\hat\sigma_{\rm gap}(T)/(\hat\sigma_B\as^{3/2})$
therefore becomes
\begin{equation}
 \begin{aligned}
 C(T)={}&\frac1{\hat\sigma_B}\int_{0<t_1<t_2<t_s<T}\!dt_1dt_2dt_s\\
 &\times\big\langle\mathcal H_{m_0}U_c(0,t_1)V_G
          U_c(t_1,t_2)V_G\Gb(t_s)\big\rangle.
 \end{aligned}
 \label{eq:sequence}
\end{equation}
The final soft insertion measures the gap veto. After integration,
a collinear insertion supplies two logarithms and a soft insertion one.
At fixed order in $\as$, replacing the former by the latter therefore
reduces the logarithmic enhancement. These contributions
enter the interplay with NGLs.

\emph{From RG evolution to a Monte Carlo algorithm.---}
The (virtual) collinear term,
$-32\pi t(C_a+C_b)\mathbf1$, is proportional to the identity in colour
space and exponentiates into a Sudakov factor. At leading colour,
$\kappa_{qq'}=16\pi\Nc$ and $\kappa_{qg}=24\pi\Nc$, giving
\begin{equation}
 \Delta_c(t_a,t_b)=e^{-\kappa_{ab}(t_b^2-t_a^2)},\qquad
 t_{\rm next}^2=t_a^2-\frac{\ln u}{\kappa_{ab}},
 \label{eq:sudakov}
\end{equation}
where $u$ is uniform on $(0,1)$.
These leading-colour Sudakov factors suffice for the leading
nonzero two-Glauber gap contribution retained here.
With $\mathcal R_c(s)$ the
real-emission contribution to $-8\pi s\Gc$, evolution with this leading
virtual term obeys
\begin{equation}
 \begin{split}
 U_c^{(0)}(t_a,t_b)={}&\Delta_c(t_a,t_b)\mathbf1\\
 &+\int_{t_a}^{t_b}ds\,\Delta_c(t_a,s)
                 \mathcal R_c(s)U_c^{(0)}(s,t_b).
 \end{split}
 \label{eq:iteration}
\end{equation}

The first term represents no emission; the second generates a collinear
emission at $s$ followed by further evolution, similarly to the purely soft
evolution in~\cite{Balsiger:2018ezi}.
Iteration therefore turns the RG evolution into the familiar
parton-shower sequence of Sudakov factors and real emissions. This
sequence evolves the colour density needed for the gap coefficient.
For $qq'$, choosing either incoming quark with probability $1/2$
reproduces the real-emission term without an additional weight. For $qg$, one can first choose the 
quark with probability $1/3$ or the
gluon with probability $2/3$, then sample its colour attachments with
signed weights. Our implementation samples the incoming leg and these
attachments jointly, as described below and in Supplemental Material,
Sec.~\ref{sec:showeralgoqg}.

We evaluate \eqref{eq:sequence} by sampling ordered insertion
times $t_1<t_2<t_s$ and generating collinear emissions before and between
the Glaubers using \eqref{eq:sudakov}. A history is one realization
of these insertion times, emissions and colour choices. It carries one
particle list and two colour flows. In $qq'$ scattering the flows remain
identical before the first Glauber. That insertion acts on one side of
the cut and creates unequal flows, which must subsequently be updated
separately for each shared emission. The emission times and particles
remain common to both sides throughout. After the second Glauber, we 
evaluate the colour trace with
$\Gb(t_s)$, integrating the soft direction over the vetoed gap. 
Sampling normalizations are given in Supplemental Material, 
Secs.~\ref{sec:showeralgoqq} and~\ref{sec:showeralgoqg}.

The leading planar Glauber phases cancel between amplitude and
conjugate amplitude. The first nonzero two-Glauber contribution is therefore
of relative order $1/\Nc^2$. To retain it, we count both explicit inverse
powers of $\Nc$ in the colour operators and the factors of $\Nc$ from
closed colour-index loops. For $qq'$, this permits projection onto equal
flows after the second Glauber; projecting earlier would remove the
required interference. For $qg$, unequal final flows can also contribute.
The rules below retain all terms at the required power using a bounded
number of relative reconnections, even at arbitrary emission multiplicity.

\emph{Quark scattering.---}
In the strict-forward geometry, particles $0,2$ point along $n_+$ and $1,3$
along $n_-$, with $n_\pm=(1,0,0,\pm1)$ and a veto in
$|\eta|<\Delta Y/2$ in the partonic centre-of-mass frame.
For the colour algebra we cross incoming quarks to outgoing antiquarks,
so an incoming quark carries an antifundamental index~\cite{Catani:1996vz}.
Writing $(i,j)$ for
a fundamental-to-antifundamental connection, the Born flows are
\begin{equation}
 \mathcal F_O=\{(2,1),(3,0)\},\qquad
 \mathcal F_S=\{(2,0),(3,1)\}.
 \label{eq:qqflows}
\end{equation}
By the Fierz identity, the exact octet tensor is proportional to
$|\mathcal F_O\rangle-|\mathcal F_S\rangle/\Nc$; the explicit
decomposition is given in Supplemental Material, Sec.~\ref{sec:showeralgoqq}.
For its leading nonzero two-Glauber contribution, the first flow suffices;
the singlet tensor already consists of a single flow. Thus each channel
starts from one diagonal flow pair, and the retained collinear evolution
uses $C_F\to\Nc/2$ and only the leading quark insertion.
After factoring out the leading power of $\Nc$ per operator,
each quark Glauber supplies $1/\Nc$, without a compensating closed
colour sum. The two insertions therefore exhaust the retained suppression:
extra inverse-colour powers from the Born tensor or collinear insertions
contribute only beyond this order (Supplemental Material,
Sec.~\ref{sec:showeralgoqq}).
These simplifications give unit collinear colour weights; the Glauber
interference must still be retained.

\begin{figure}[t]
 \centering
 \includegraphics[width=0.48\textwidth]{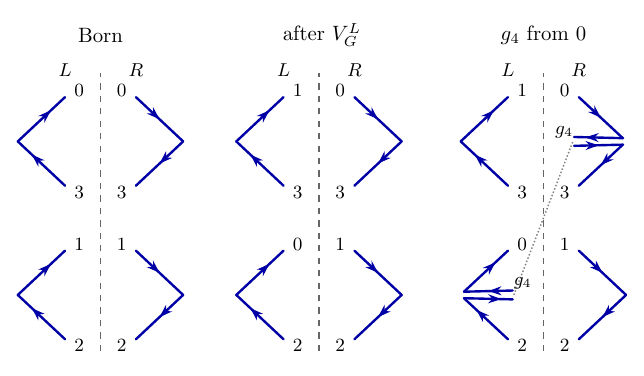}
 \caption{An octet colour pair at Born level, after a Glauber insertion
 on the amplitude, and after emission of $g_4$ from incoming quark $0$.
 Dashed lines separate amplitude ($L$) and conjugate amplitude ($R$).
 Arrows follow fundamental indices, reversing on conjugation; they do not
 indicate momenta. The dotted connector identifies the same gluon $g_4$
 on both sides, inserted into $(2,0)$ on $L$ and $(3,0)$ on $R$.}
 \label{fig:flow-history}
\end{figure}

The singlet and octet channels use the same evolution and measurement
rules; only these initial colour connections differ. At each collinear
emission, we choose the incoming quark $b=0$ or $1$ with equal probability
and insert a shared gluon $g$ by splitting
$(a,b)\to(a,g)+(g,b)$ in each flow. Its colour neighbours may differ
between amplitude and conjugate amplitude (Fig.~\ref{fig:flow-history}).

The Fierz identity gives $V_G=-4i\pi(S_L-S_R)$, where $S$ swaps the original
incoming anti-colour ends. Each of the two insertions acts on either
$L$ or $R$, giving the combinations $LL,LR,RL,RR$. Summing these terms
gives signs $s_{LL}=s_{RR}=-1$ and $s_{LR}=s_{RL}=+1$, with common factor
$16\pi^2$. After the second insertion, unequal flow pairs lose at least two
colour loops and are beyond the retained accuracy. Equal pairs survive with
unit planar trace. This projection is applied only at the end: the first
Glauber creates precisely the interference that the second must act on.

The terminal soft insertion measures radiation into the gap.
Its angular dependence is given by the integral of the soft radiator
$W_{ij}^{\,k}=n_i\!\cdot n_j/[(n_i\!\cdot n_k)(n_j\!\cdot n_k)]$
over the gap with measure $d\Omega_k/(4\pi)$. The colour charges and
overall normalization are specified in \eqref{eq:soft-gap}.
The radiator vanishes when the two lightlike directions coincide. For opposite
beam directions its integral over the gap is $\Delta Y$. A surviving
planar colour connection contributes $\bm T_i\!\cdot\bm T_j=-\Nc/2$,
so the gap trace is $-4\Nc\Delta Y D$, where $D$ counts connections
between opposite directions (see Supplemental Material, Sec.~\ref{app:BNSqq}). 
In units of $4\Nc\Delta Y$, the resulting colour
weight is $Z=-\sum_{\lambda}s_\lambda P_\lambda D_\lambda$.
Here $\lambda\in\{LL,LR,RL,RR\}$ labels these four choices of sides,
and $P_\lambda$ is one for equal final flows and zero otherwise.
Projection and measurement are distinct: the Born singlet has $P=1$
but $D=0$.

\emph{An incoming gluon.---}
For $qg\to qV$, incoming quark $0$ and outgoing quark $2$ have the
same direction, opposite to incoming gluon $1$. We describe the colour
density by pairs of amplitude and conjugate-amplitude flows, as in
CVolver~\cite{DeAngelis:2020rvq}. The traceless Born tensor is
\begin{equation}
 |B\rangle=|\{(1,0),(2,1)\}\rangle
 -\frac1{\Nc}|\{(1,1),(2,0)\}\rangle .
 \label{eq:qgborn}
\end{equation}
Its density-matrix  $\rho_B=|B\rangle\langle B|$ obeys
$V_G\rho_B=0$: at least one collinear
emission must precede the first Glauber. On each side of the cut the
real-emission maps are
\begin{equation}
 B_q=-I_{\bar0}+\frac{L_g}{\Nc},\qquad
 B_g=I_{1_3}-I_{1_{\bar3}},
 \label{eq:realqg}
\end{equation}
where $I_e$ inserts the new gluon at end $e$ and $L_g$ appends a
disconnected trace; $3,\bar3$ denote fundamental and antifundamental ends.
The normalized density map is
$\mathcal R\rho=(B_q\rho B_q^\dagger+B_g\rho B_g^\dagger)/(3\Nc)$.
Each beam gives four terms, combining two attachments on each side.
For the gluon these include negative mixed-end interferences, so unequal
flows arise even before the first Glauber. Each term adds one shared
particle and one connection per flow.

Emission times are generated by the Sudakov in \eqref{eq:sudakov}.
Each history carries a signed colour weight $W$, initially one.
At each emission we expand $\mathcal R\rho$, discard terms beyond the
retained colour order and combine identical outcomes. Each flow pair and
its transpose form one sampling alternative, with signed coefficient
$a_j$. These coefficients are real after removing the common Glauber
phases; explicit inverse powers of $\Nc$ are tracked separately.
We select alternative $j$ with probability $p_j$ and compensate for this
choice by updating the colour weight:
\begin{equation}
 p_j=\frac{|a_j|}{\sum_k|a_k|},\qquad
 W\longmapsto W\frac{a_j}{p_j}.
 \label{eq:signed}
\end{equation}
The weighted average recovers the local sum. The beam and its
attachments are selected jointly; the leading Casimir ratio $1:2$ does
not prescribe their sampling frequencies. The first real image of the
Born density is kept exact until the next operation. Subsequently, only
one alternative and its weight are propagated.

The incoming charge product is $Q\equiv2\bm T_0\cdot\bm T_1=P-K$.
$P$ swaps the incoming anti-colour ends; $K$ joins gluon colour to quark
anti-colour and reconnects their former partners. If these ends are
already connected, $K$ supplies a closed colour sum $\Nc$. Thus
\begin{equation}
 G=\frac{P_L-K_L-P_R+K_R}{\Nc},\qquad V_G=-4i\pi\Nc G,
 \label{eq:qgglauber}
\end{equation}

This is the colour-flow decomposition of the incoming-charge
commutator. When the first Glauber is sampled, its four terms act on
each amplitude/conjugate-amplitude pair in the current state.
After the colour truncation and combination of identical outcomes,
one resulting alternative is selected with \eqref{eq:signed}.
To organize colour accuracy, write a term with $M$ connections as
$W\Nc^{2-M-p}|f_L\rangle\langle f_R|$. Its trace divided by $\Nc^2$
is $W\Nc^{-h}$, where
\begin{equation}
 h=d+p,\qquad d=M-\operatorname{cycles}(f_R^{-1}f_L).
 \label{eq:grade}
\end{equation} 
The flows are bijections between ordered sets of fundamental and
antifundamental endpoints; the cycles of $f_R^{-1}f_L$ count closed
index loops.
Combining explicit inverse-colour powers with flow-overlap
suppression also underlies CVolver's colour truncation~\cite{Forshaw:2025bmo}.
Here we retain $h\leq2$, which contains the leading nonzero two-Glauber
contribution. The current connections and explicit power $p$
determine $h$ before the final soft contraction.
Each normalized local map changes $h$ by zero or two:
a Glauber can reduce $d$, but its explicit $1/\Nc$ compensates that
reduction. A term discarded at $h>2$ can therefore never contribute
to the retained $h\leq2$ trace at a later step.
Since $p\geq0$, retained pairs have $d\leq2$: one flow can be obtained
from the other by at most two partner swaps. We store one complete
connection list and these swaps; the list itself grows with every emission.

Unlike $qq'$, unequal final pairs and real emissions between the two
Glauber exchanges contribute. Finally, we apply the soft operator 
integrated over the gap and close the colour trace, as detailed 
in Supplemental Material, Sec.~\ref{sec:showeralgoqg}.

CVolver includes Coulomb exchanges in its soft anomalous dimension,
regulated by a collinear cutoff~\cite{Forshaw:2025fif}.
Our algorithm uses the renormalized collinear anomalous dimension without an auxiliary angular cutoff.
A direct mapping between these formulations remains to be established
and could help to guide extensions combining collinear and wide-angle radiation.

\begin{figure*}[t]
 \centering
 \includegraphics[width=\textwidth]{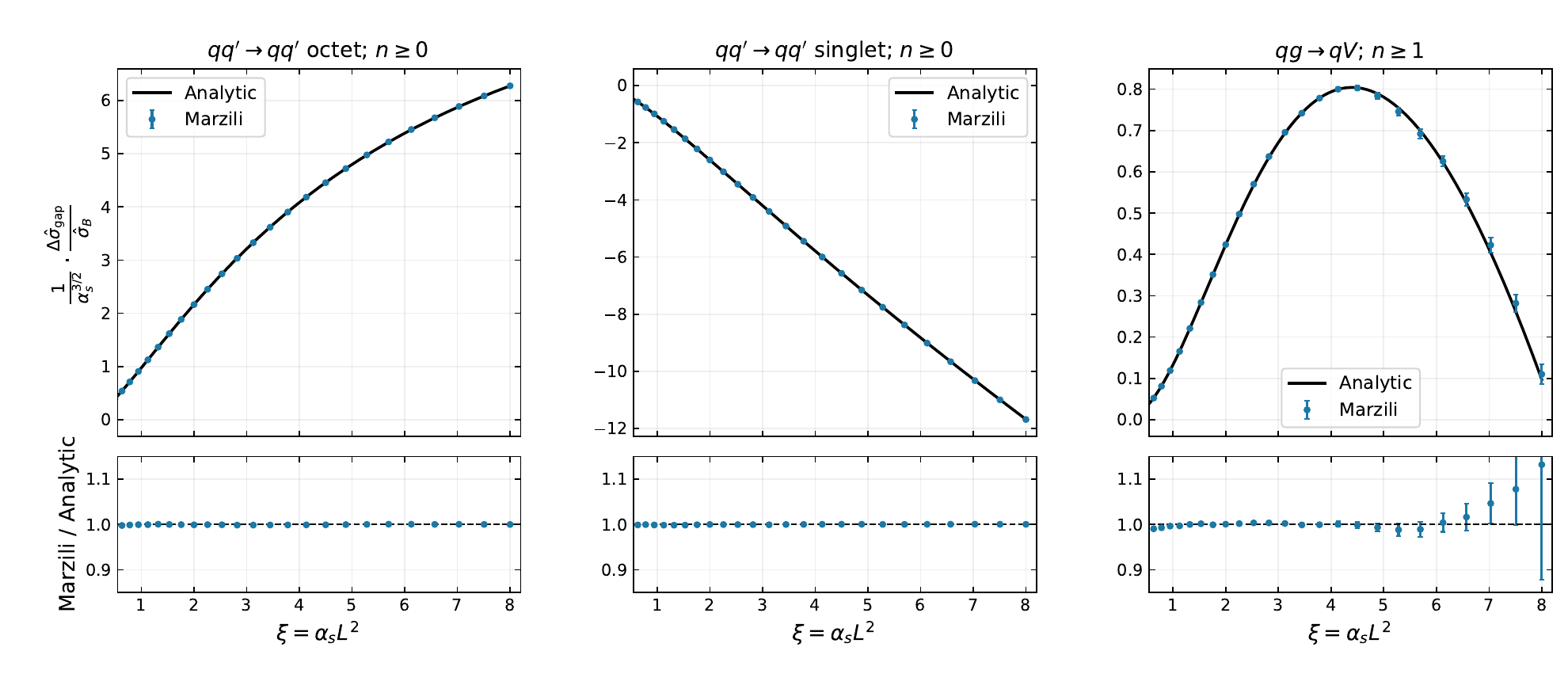}
 \caption{All-order comparison of the Marzili calculation (points) with
 analytic resummations (curves, taken from \cite{Becher:2021zkk,Becher:2023mtx}) 
 for forward $qq'\to qq'$ in the octet
 and singlet channels; and $qg\to qV$.
 The upper panels show the Born-normalized gap correction
 $C=\Delta\hat\sigma_{\rm gap}/(\hat\sigma_B\as^{3/2})$
 at relative order $1/\Nc^2$ in the colour expansion, with $\Nc=3$
 and full gap width $\Delta Y=2$. The $qq'$ results include the
 three-loop ($n=0$) term; the $qg$ result starts at four loops.
 Lower panels show Marzili/analytic ratios. Error bars denote the Monte
 Carlo error.}
 \label{fig:comparison}
\end{figure*}

\emph{All-order results.---}
For $qq'$, a real emission between the two Glaubers gives zero after
the final colour projection. Summing over the generated histories
therefore supplies the no-emission factor $D_G=e^{-\kappa_{qq'}(t_2^2-t_1^2)}$.
Before the first Glauber, each beam independently has no emission with probability
$a=e^{-\kappa_{qq'}t_1^2/2}$. In the octet channel, zero, one or two beams
having radiated gives colour weights $+4,0,-4$, respectively, when the
middle interval is empty. Their probabilities $a^2$, $2a(1-a)$ and
$(1-a)^2$ give the average weight $4(2a-1)$. The singlet weight is $-4$
for every such history. Integrating the terminal time gives $T-t_2$, so
\begin{equation}
 \begin{aligned}
 \begin{pmatrix}C_O(T)\\C_S(T)\end{pmatrix}
 ={}&256\pi^2\Nc\Delta Y\int_0^Tdt_2\,(T-t_2)\\
 &\times\int_0^{t_2}dt_1\,D_G
       \begin{pmatrix}2a-1\\-1\end{pmatrix}.
 \end{aligned}
 \label{eq:probabilities}
\end{equation}
This makes the connection to \cite{Boer:2024xzy} explicit:
its fixed-coupling Sudakov factors obey
$a=U_c(1/2;Q,\mu_1)$ and $D_G=U_c(1;\mu_1,\mu_2)$, with $\mu_i$
related to $t_i$ by \eqref{eq:variables}. Equation~\eqref{eq:probabilities}
is the two-Glauber contribution to its large-$\Nc$ quark resummation,
projected onto the forward octet and singlet Born states. Evaluating the
integrals gives the benchmark functions of Ref.~\cite{Becher:2021zkk},
as shown in Supplemental Material, Sec.~\ref{app:BNSqq}.
For $qg$, real emissions between the Glaubers also contribute.

Figure~\ref{fig:comparison} demonstrates agreement with the independent
all-order resummations in all three channels, including
the pronounced $qg$ turnover. Its shape reflects cancellations
between contributions with different collinear evolution factors.
This validates the numerical resummation of the fixed-coupling
two-Glauber tower. Reference formulas and
numerical details are given in Supplemental Material,
Secs.~\ref{app:BNSqq}, \ref{app:BNSSqg} and~\ref{sec:numerics}.

We reduce fluctuations by combining cancelling colour terms within
each event. For $qq'$, the $LL,RR,LR,RL$ contributions are summed for the
same collinear history. For $qg$, terms producing identical reconnections
in Eq.~\eqref{eq:qgglauber} are combined before sampling, amplitude and
conjugate-amplitude terms are kept correlated, and the final Glauber and
gap contraction are summed exactly. These local cancellations improve
efficiency, but sizeable cancellations between positive and negative event
weights remain, particularly in $qg$.

\emph{Conclusions.---}
We have achieved the first Monte Carlo resummation of SLLs
involving two Glauber exchanges by solving RG evolution equations
in Marzili. For forward
quark and quark--gluon scattering, our algorithm reproduces the known 
all-order results at relative order $1/\Nc^2$ in the colour
expansion. Marzili now provides separate resummations of NGLs and
SLLs series, an important milestone towards studying their interplay
in a combined evolution.

An immediate extension is to include successive wide-angle soft
emissions and their virtual corrections alongside the collinear radiation
and Glauber exchanges. The simpler colour structure makes quark scattering
a natural first setting. Other incoming channels and additional Glauber
exchanges offer further extensions. The present calculation shows
that an iterative Monte Carlo evolution can retain the Glauber interference
needed to resum SLLs, opening a path towards their inclusion in parton
showers.

\emph{Note added.---} While completing this work, we became aware of related progress
developing a parton shower framework, also capable of capturing SLLs \cite{JackUpcoming}.

\emph{Acknowledgements.---}
We are particularly grateful to Thomas Becher for his sustained
support and many valuable discussions throughout this work. We thank
Dominik Schwienbacher and Patrick Hager for comments
on the manuscript, and Jack Helliwell and Xiaofeng Xu for stimulating
discussions about colour evolution.
Nicolas Schalch is funded through the Royal Society,
grant RP\textbackslash R\textbackslash 231001.

\bibliographystyle{letter-apsrev4-2}
\bibliography{letterbib}

\clearpage
\onecolumngrid
\appendix
\setcounter{secnumdepth}{1}
\setcounter{page}{1}
\hypersetup{pageanchor=false}
\section*{Supplemental material}
\numberwithin{equation}{section}

This supplement gives the analytic reference formulas and the evolution
algorithms. We resum arbitrary incoming-collinear radiation with two
Glauber exchanges and one final soft insertion integrated over the gap.
We keep the first nonvanishing term in the colour expansion,
of relative order $1/\Nc^2$. All expressions use fixed coupling and the
strict-forward geometry, in which every coloured particle lies along one
of the two beam directions. The full rapidity-gap width is $\Delta Y$.

\section{Soft anomalous dimension and analytic quark benchmarks}\label{app:BNSqq}

The purely soft anomalous dimension used in Eq.~\eqref{eq:generator} is the
combination of real and virtual contributions of Refs.~\cite{Becher:2021zkk,Becher:2023mtx}.
For clarity, it can be written as
\begin{equation}
 \Gb=2\sum_{i\ne j}(\bm T_{i,L}\cdot\bm T_{j,L}
                  +\bm T_{i,R}\cdot\bm T_{j,R})
       \int\frac{d\Omega_k}{4\pi}\overline W_{ij}^{\,k}
 -4\sum_{i\ne j}\int\frac{d\Omega_k}{4\pi}\,
       \bm T_{i,L}\circ\bm T_{j,R}\,
       \overline W_{ij}^{\,k}\Theta_{\rm allow}(n_k).
 \label{eq:soft-full}
\end{equation}
The sums run over ordered pairs. The real term creates an additional colour index
and direction before subsequent operators act. Here
$W_{ij}^{\,k}=n_i\cdot n_j/[(n_i\cdot n_k)(n_j\cdot n_k)]$ and the bar
subtracts its collinear limits, as in Eq.~(9) of Ref.~\cite{Becher:2021zkk}. 
The separate PDF anomalous dimension is not part of
Eq.~\eqref{eq:soft-full}. Under a terminal inclusive colour trace, real and
virtual contributions cancel in the allowed region. Since the gap excludes
every hard and beam-collinear direction, the collinear subtractions have no
support there. Consequently
\begin{equation}
 \langle H\Gb\rangle=\Tr\!\left[H\,\mathcal S_{\rm gap}\right],
 \qquad
 \mathcal S_{\rm gap}=8\sum_{i<j}\bm T_i\cdot\bm T_j
       \int_{\rm gap}\frac{d\Omega_k}{4\pi}W_{ij}^{\,k}.
 \label{eq:soft-gap}
\end{equation}
In the forward limit the integral is $\Delta Y$ for opposite directions and
zero for equal directions. In particular
$W_{+-}=2/\sin^2\theta$ and
$d\Omega_k\,W_{+-}/(4\pi)=d\eta\,d\phi/(2\pi)$.
The negative gap contribution for a planar colour dipole follows from
$\bm T_i\cdot\bm T_j\to-\Nc/2$.

The scale integrals multiplying the colour coefficients $C_{rn}$
in Eq.~\eqref{eq:words}
are given by Eq.~(14) of Ref.~\cite{Becher:2021zkk}:
\begin{equation}
 \frac{\Delta\hat\sigma_n}{\hat\sigma_B}
 =\left(\frac{\as}{4\pi}\right)^{n+3}L^{2n+3}
   \frac{(-4)^n n!}{(2n+3)!}
   \sum_{r=0}^n\frac{(2r)!}{4^r(r!)^2}\,
                         \frac{C_{rn}}{\hat\sigma_B}.
 \label{eq:scale-integrals}
\end{equation}

To match the horizontal axis of Fig.~\ref{fig:comparison}, we write the
same cumulative coefficient as $C(\xi)\equiv C(T=\sqrt\xi/(4\pi))$.
Summing Eq.~\eqref{eq:scale-integrals} over $n$, dividing by
$\as^{3/2}$ and using $\xi=\as L^2$ gives the plotted quantity.
Here $n$ counts collinear operators, including virtual ones, rather
than emitted particles:
\begin{equation}
 \begin{aligned}
 C(\xi)&=\frac1{\as^{3/2}}\sum_{n=0}^{\infty}
                 \frac{\Delta\hat\sigma_n}{\hat\sigma_B}\\
 &=\frac{\xi^{3/2}}{(4\pi)^3}\sum_{n=0}^{\infty}
 \left(-\frac\xi\pi\right)^n\frac{n!}{(2n+3)!}
 \sum_{r=0}^n\frac{(2r)!}{4^r(r!)^2}
                      \frac{C_{rn}}{\hat\sigma_B}.
 \end{aligned}
 \label{eq:plotted-coefficient}
\end{equation}
The $C_{rn}$ in this equation are expanded to the retained colour
order. The Monte Carlo points estimate the same coefficient through
Eqs.~\eqref{eq:qq-estimator} and~\eqref{eq:qg-estimator}.
Changing the cumulative endpoint from $T$ to $\xi$ introduces no Jacobian;
we are not plotting a differential distribution.

We retain the reference labels on the angular integrals
$J_j$~\cite{Becher:2021zkk,Becher:2023mtx}: incoming legs $1,2$
there correspond to $0,1$ here, and outgoing legs $3,4$ to $2,3$.
In our radiator labels their definitions are
\begin{equation}
 \begin{aligned}
 J_{12}=J_2&=\int_{\rm gap}\frac{d\Omega_k}{4\pi}W_{01}^{\,k},\\
 J_j&=\int_{\rm gap}\frac{d\Omega_k}{4\pi}
       \bigl(W_{0,j-1}^{\,k}-W_{1,j-1}^{\,k}\bigr),\quad j=3,4,
 \qquad J_{43}=J_4-J_3.
 \end{aligned}
 \label{eq:reference-angular-integrals}
\end{equation}
Equal-direction radiators vanish, while opposite-direction
integrals give $\Delta Y$. Thus $J_2=\Delta Y$, $J_3=-\Delta Y$,
$J_4=\Delta Y$ and $J_{43}=2\Delta Y$.
With these values, the octet bracket in Eq.~(19) of
Ref.~\cite{Becher:2021zkk} contains
$C_FJ_{43}\to\Nc\Delta Y$ and
$J_2(\Nc^2-2^{r+1}+1)/\Nc\to\Nc\Delta Y$.
The singlet bracket reduces to $-2C_F\Delta Y\delta_{r0}$.
Keeping the highest power of $\Nc$ at each fixed $n,r$ therefore gives
\begin{equation}
 \frac{C^{O}_{rn}}{\hat\sigma_B}
 =2^{8-r}\pi^2(4\Nc)^n\Nc\Delta Y(2-\delta_{r0}),\qquad
 \frac{C^{S}_{rn}}{\hat\sigma_B}
 =-2^{8-r}\pi^2(4\Nc)^n\Nc\Delta Y\,\delta_{r0}.
 \label{eq:qqwords}
\end{equation}
The expansion is performed before setting $\Nc=3$. Subleading pieces of
$C_F$ must not be reinstated in an otherwise truncated result.

In terms of the plotted variable, define the common prefactor and
resummation variable
\begin{equation}
 P(\xi)=\frac{2\Nc\Delta Y}{3\pi}\xi^{3/2}
             =\frac{128\pi^2\Nc\Delta Y}{3}T^3,\qquad
 w=\frac{\Nc}{\pi}\xi=16\pi\Nc T^2.
 \label{eq:prefactor}
\end{equation}
We use the regular kernel
\begin{equation}
 H(z)=6\int_0^1dy\,y(1-y)e^{-zy^2}
     =\frac3z-\frac{3\sqrt\pi}{2z^{3/2}}\operatorname{erf}(\sqrt z),
 \qquad H(0)=1.
 \label{eq:H}
\end{equation}
Summing Eq.~\eqref{eq:scale-integrals} gives
\begin{align}
 C_O(\xi)&=P(\xi)F_O(w),& C_S(\xi)&=-P(\xi)F_S(w),\label{eq:qq-reference}\\
 F_O(w)&=\int_0^1du\,
       \bigl[2H(w(1-u^2/2))-H(w(1-u^2))\bigr],&
 F_S(w)&=\int_0^1du\,H(w(1-u^2)).\label{eq:qq-functions}
\end{align}
These functions also follow directly from
Eq.~\eqref{eq:probabilities}: setting $t_2=Ty$ and $t_1=t_2u$ gives
$D_G=e^{-wy^2(1-u^2)}$ and $aD_G=e^{-wy^2(1-u^2/2)}$.
The remaining $y$ integration is the kernel $H$ in Eq.~\eqref{eq:H},
with the normalization $P(\xi)$ in Eq.~\eqref{eq:prefactor}.

For comparison with the scalar evolution factors in Eq.~(1.5) of
Ref.~\cite{Boer:2024xzy}, fixed coupling and $\mu_h=Q$ give
$U_c(v;\mu_i,\mu_j)=\exp[-v\kappa_{qq'}(t_j^2-t_i^2)]$.
Thus $D_G=U_c(1;\mu_1,\mu_2)$ and
$aD_G=U_c(1/2;Q,\mu_1)U_c(1;\mu_1,\mu_2)$ are precisely the evolution
factors entering its Eq.~(2.14) for two Glauber insertions.
The comparison uses the same forward Born colour states and sets
$2C_F/\Nc\to1$ consistently with the colour expansion.
The benchmark functions have expansions
\begin{equation}
 F_O(w)=1-\frac3{10}w+\frac9{140}w^2+O(w^3),\qquad
 F_S(w)=1-\frac15w+\frac4{105}w^2+O(w^3).
 \label{eq:qq-smallw}
\end{equation}
Figure~\ref{fig:comparison} uses the full functions, including the
three-loop $n=0$ term. The SLL-only sum starting at $n=1$, as plotted in
Ref.~\cite{Becher:2021zkk}, instead uses $F_{O,S}\to F_{O,S}-1$. In physical normalization,
\begin{equation}
 \begin{aligned}
 \frac{\Delta\hat\sigma_O}{\hat\sigma_B}
 &=\Delta Y\left[\frac{2\Nc}{3\pi}\as^3L^3
                 -\frac{\Nc^2}{5\pi^2}\as^4L^5+\cdots\right],\\
 \frac{\Delta\hat\sigma_S}{\hat\sigma_B}
 &=\Delta Y\left[-\frac{2\Nc}{3\pi}\as^3L^3
                 +\frac{2\Nc^2}{15\pi^2}\as^4L^5+\cdots\right].
 \end{aligned}
 \label{eq:qq-physical}
\end{equation}
The numerical benchmarks use the full functions, with a series for $H(z)-1$
near the origin to avoid cancellation. For example, at $\Nc=3$,
$\Delta Y=2$ and $\xi=8$, these formulas give
$C_O=6.2754011$ and $C_S=-11.6804371$ in the plotted normalization.

\section{Analytic quark--gluon benchmark}\label{app:BNSSqg}

The forward $qg\to qV$ benchmark is the leading nonzero colour term of
Eqs.~(7.16) and (7.17) of Ref.~\cite{Becher:2023mtx}.
The same definitions and label translation in
Eq.~\eqref{eq:reference-angular-integrals} give $J_{12}=\Delta Y$
and $J_3=-\Delta Y$: the source's outgoing leg $3$ is our quark $2$,
parallel to incoming quark $0$.
A useful intermediate step in expanding Eq.~(7.16) is
\[
 \frac{\Nc(\Nc+3)}{2(\Nc+1)}(3\Nc+2)^r
 -\frac{\Nc(\Nc-3)}{2(\Nc-1)}(3\Nc-2)^r
 =(3\Nc)^r\left[2\left(1+\frac r3\right)
                  +O(\Nc^{-2})\right].
\]
The difference must be formed before expanding away the eigenvalue
splitting: it produces the term proportional to $r$. Combining the
remaining terms in the forward geometry then gives
\begin{equation}
 \frac{C^{qg}_{rn}}{\hat\sigma_B}
 =-256\pi^2(4\Nc)^n\Nc\Delta Y(1-\delta_{r0})
 \left[2\left(1+\frac r3\right)\left(\frac32\right)^r
       -4\left(\frac12\right)^r\right].
 \label{eq:qgwords}
\end{equation}
In particular, $C^{qg}_{0n}=0$ for all $n$, and
$C^{qg}_{11}/\hat\sigma_B=-2048\pi^2\Nc^2\Delta Y$.
The vanishing follows already from
$Q|B\rangle=-\Nc|B\rangle$, so $[Q,|B\rangle\langle B|]=0$:
collinear evolution before the first Glauber is necessary.

Using the resummation function in Eq.~(5.23) of Ref.~\cite{Becher:2023mtx}, write
\begin{equation}
 \Sigma(v,w)=\int_0^1du\,H\!\left(w[1+(v-1)u^2]\right),\qquad
 C_{qg}(\xi)=-P(\xi)B(w),
 \label{eq:sigma}
\end{equation}
where $\Sigma(v,w)$ is an auxiliary function, not a cross section.
Its integral follows from Eq.~(5.23) by setting the integration
variable there to $u^2$. The overall factor in Eq.~(7.17), after division
by $\hat\sigma_B\as^{3/2}$, is $-P(\xi)$. Expanding its first line in
$1/\Nc$ at fixed $w=\Nc\as L^2/\pi$ gives
\begin{equation}
 B(w)=2\Sigma(0,w)-4\Sigma(\tfrac12,w)+2\Sigma(\tfrac32,w)
       +\left.\partial_v\Sigma(v,w)\right|_{v=3/2}.
 \label{eq:qg-reference}
\end{equation}
The derivative acts on the first argument at fixed $w$ and is
essential. In the general-$\Nc$ expression, the two
arguments $v_\pm=3/2\pm1/\Nc$ have coefficients
\begin{equation}
 A_\pm=\frac{\Nc(\Nc\pm3)}{2(\Nc\pm1)}
       =\frac{\Nc}{2}\pm1+O(\Nc^{-1}),\qquad
 A_+\Sigma(v_+,w)-A_-\Sigma(v_-,w)
 \longrightarrow 2\Sigma(\tfrac32,w)+\partial_v\Sigma(\tfrac32,w).
 \label{eq:qg-derivative}
\end{equation}
The remaining coefficient
$(4\Nc^2-2)/(\Nc^2-1)$ tends to $4$, while $A_+-A_-\to2$.
These give the coefficients $2$ of $\Sigma(0,w)$ and $-4$ of
$\Sigma(1/2,w)$ in Eq.~\eqref{eq:qg-reference}.
Setting the arguments equal before expanding would miss the derivative.
Likewise, the last line of Eq.~(7.17) in Ref.~\cite{Becher:2023mtx},
already evaluated at $\Nc=3$,
is not the colour expansion used here.

The derivative has a direct counterpart in the large-$\Nc$ evolution
of Ref.~\cite{Boer:2024xzy}. With its scalar factors, define
\begin{equation}
 K_v(t_1,t_2)=U_c(v;Q,\mu_1)U_c(1;\mu_1,\mu_2)
 =e^{-\kappa_{qq'}[t_2^2+(v-1)t_1^2]}.
 \label{eq:qg-rg-kernel}
\end{equation}
At fixed coupling, the same ordered integrals give
\begin{equation}
 \begin{aligned}
 \Sigma(v,w)&=\frac6{T^3}\int_0^Tdt_2\,(T-t_2)
                    \int_0^{t_2}dt_1\,K_v(t_1,t_2),\\
 \partial_v\Sigma(v,w)&=\frac6{T^3}\int_0^Tdt_2\,(T-t_2)
                    \int_0^{t_2}dt_1\,I_h(Q,\mu_1)K_v(t_1,t_2),
 \qquad I_h(Q,\mu_1)=-\kappa_{qq'}t_1^2.
 \end{aligned}
 \label{eq:qg-rg-sigma}
\end{equation}
Here $I_h$ is the logarithmic integral in Eq.~(3.5) of that reference.
Its Eq.~(3.4) shows how the coalescing eigenvalues
$v_\pm=3/2\pm1/\Nc$ produce the factor $I_h U_c(3/2)$; this is precisely
the origin of $\partial_v\Sigma(3/2,w)$ in
Eq.~\eqref{eq:qg-reference}. Thus the benchmark is the fixed-coupling,
forward $qg\to qV$ projection of the contribution with two Glauber
insertions in that evolution. The middle factor $U_c(1;\mu_1,\mu_2)$
includes the effect of real and virtual collinear evolution. It is not
the $qg$ no-emission probability, which has eigenvalue $3/2$;
real emissions between the Glaubers must therefore be retained.

A stable integral representation is
\begin{equation}
 \begin{split}
 B(w)=\int_0^1du\,\bigl\{
 &2[H(w(1-u^2))-1]-4[H(w(1-u^2/2))-1]\\
 &+2[H(w(1+u^2/2))-1]+wu^2H'(w(1+u^2/2))\bigr\}.
 \end{split}
 \label{eq:qg-integral}
\end{equation}
Its small-$w$ expansion and first physical contribution are
\begin{equation}
 B(w)=-\frac w5+\frac{11w^2}{84}-\frac{w^3}{21}+O(w^4),\qquad
 \frac{\Delta\hat\sigma_{qg}}{\hat\sigma_B}
 =\frac{2\Nc^2\Delta Y}{15\pi^2}\as^4L^5+O(\as^5L^7).
 \label{eq:qg-smallw}
\end{equation}
There is no $n=0$ contribution. The quadrature evaluates $H-1$ and $H'$ by
series near zero. Equations~\eqref{eq:qq-functions} and
\eqref{eq:qg-integral} specify the comparison curves independently of the
stochastic evolution. 

Explicitly, the plotted curve is
$C_{qg}(\xi)=-P(\xi)B(\Nc\xi/\pi)$, with $P(\xi)$ defined in
Eq.~\eqref{eq:prefactor} and $B$ evaluated using
Eq.~\eqref{eq:qg-integral}.
At $\Nc=3$, $\Delta Y=2$ and $\xi=8$, the
quark--gluon curve has $C_{qg}=0.0970510$, with the same normalization as
the quark curves and the Marzili results in Fig.~\ref{fig:comparison}.

\section{Monte Carlo algorithm for $qq'\to qq'$}\label{sec:showeralgoqq}

This section specifies how one event is generated and converted into
a contribution to the cumulative coefficient $C(T)$. The event contains
one shared particle list and separate amplitude and conjugate-amplitude
connection lists. All directions are $n_+$ or $n_-$; energies and recoil
are absent in the collinear approximation. Initially both flows equal
the chosen Born flow in Eq.~\eqref{eq:qqflows}.
For octet exchange, the exact tensor follows from
\[
 t^a_{i_2i_0}t^a_{i_3i_1}
 =\frac12\left(\delta_{i_2i_1}\delta_{i_3i_0}
 -\frac1{\Nc}\delta_{i_2i_0}\delta_{i_3i_1}\right).
\]
The two products of Kronecker deltas are precisely
$\mathcal F_O$ and $\mathcal F_S$, respectively.

The Born and collinear simplifications follow from the counting in
Eq.~\eqref{eq:grade}. After factoring out the leading $\Nc$ per Glauber,
the normalized quark map is $(S_L-S_R)/\Nc$.
Each insertion increases the explicit inverse-colour power $p$ by one,
so two Glaubers necessarily give $p\geq2$. The subleading Born terms,
normalization correction and quark trace insertions add further inverse
powers, as does the subleading scalar virtual term. Neither later
insertions nor the normalized soft contraction can lower $p$.
Since the loop deficit is nonnegative, these extra powers cannot
contribute at $h=p+d=2$. The retained histories therefore have $p=2,d=0$,
which also justifies the final equal-flow projection.

\emph{Generating collinear emissions.}
Let $E_i$ insert the same new gluon next to incoming quark $i$ in
both flows, splitting the connection ending on that quark separately
on each side of the cut. At this colour accuracy, the collinear anomalous dimension is
\begin{equation}
 \Gc=4\Nc\left[\mathbf1-\frac12(E_0+E_1)\right],\qquad
 -8\pi t\Gc=\lambda_{qq'}(t)
 \left[\frac12(E_0+E_1)-\mathbf1\right],\qquad
 \lambda_{qq'}(t)=2\kappa_{qq'}t.
 \label{eq:qq-jump}
\end{equation}
Starting at $t_a$, draw a uniform number $u$ and obtain the next
emission time by Sudakov inversion, Eq.~\eqref{eq:sudakov}.
If it lies beyond the interval endpoint $t_b$, there is no further
emission there. Otherwise choose quark $0$ or $1$ with probability
$1/2$, apply $E_i$, and repeat from the new time.
The probabilities for no emission and for a first emission anywhere
in the interval sum to one:
\begin{equation}
 \Delta_c(t_a,t_b)+\int_{t_a}^{t_b}ds\,
             \Delta_c(t_a,s)\lambda_{qq'}(s)=1.
 \label{eq:qq-unitarity}
\end{equation}
Thus the Sudakov is already included in the generated histories;
no additional no-emission or collinear colour weight is required.

\emph{One complete event.}
\begin{enumerate}
 \item Draw three independent uniform numbers on $(0,T_{\max})$
 and sort them to obtain $t_1<t_2<t_s$. These are the two Glauber
 insertion times and the terminal soft time. Their joint density is
 $6/T_{\max}^3$.
 \item Generate emissions between $0$ and $t_1$ as above, updating
 the shared particles and both flows.
 \item Apply the first Glauber swap on the left or on the right
 as two algebraic alternatives. Generate one common set of emission
 times and beam choices between $t_1$ and $t_2$, and apply these emissions
 to both alternatives. Their connections can differ, but they describe
 the same emitted particles.
 \item Apply the second swap on either side, giving the four terms
 $LL,LR,RL,RR$, with signs $s_{LL}=s_{RR}=-1$ and
 $s_{LR}=s_{RL}=+1$. For each term $\lambda$, set $P_\lambda=1$
 if the final flows are equal and $P_\lambda=0$ otherwise.
 For a surviving common flow, count its opposite-direction connections
 $D_\lambda$.
 \item Form $Z_e=-\sum_\lambda s_\lambda P_\lambda D_\lambda$
 and add the contribution below to every endpoint $T\geq t_s$.
\end{enumerate}
\begin{equation}
 X_e(T)=\frac{T_{\max}^3}{6}\,64\pi^2\Nc\Delta Y\,
                   Z_e\,\Theta(T-t_s).
 \label{eq:qq-estimator}
\end{equation}
Its sample mean estimates $C(T)$, with $\as^{3/2}$ removed.
The inverse time density is $T_{\max}^3/6$; $64\pi^2\Nc\Delta Y$
combines the two Glauber factors with the angle-integrated gap
measurement. Their relative signs are already included in $Z_e$.

\emph{Final colour contraction.}
Each closed index loop supplies $\Nc$. Equal flows with $M$
connections give $M$ loops. After two quark Glauber swaps, an unequal
pair differs by a nonidentity even permutation and loses at least two
loops. A soft reconnection can recover at most one loop while losing
the explicit $\Nc$ of a diagonal antenna, so the term remains beyond
the retained order. This argument applies after the second Glauber:
the mismatch created by the first can still be undone by the second
and must be retained during the intervening evolution.

After the first Glauber, let $a\ne b$ be the colour neighbours
of incoming end $0$ in the two flows. A first middle emission $g$ from
that end replaces $(a,0)$ and $(b,0)$ by $(a,g),(g,0)$ and
$(b,g),(g,0)$, respectively. The mismatch is now at the emitted
anticolour end $g$. Later collinear emissions and the second Glauber
act only at incoming ends $0,1$, so they cannot repair it.
The same argument holds for emission from end $1$ and for either side
of the first Glauber. Thus any real emission between the Glaubers
prevents final equality and gives $Z_e=0$.

The calculation generates these histories explicitly; their sum is
equivalent to keeping an empty interval with probability
$\Delta_c(t_1,t_2)$. For an empty interval,
$Z=2[D(f)-D(Sf)]$, where $f$ is the flow at $t_1$ and $S$ swaps its
incoming ends. In the octet channel the weights are $+4,0,-4$ for
zero, one or two beams having radiated before $t_1$, respectively,
giving Eq.~\eqref{eq:probabilities}. In the singlet channel
$D(f)=0$, $D(Sf)=2$ and $Z=-4$.
Replacing the gap measurement by the inclusive planar trace gives
$\sum_\lambda s_\lambda P_\lambda=0$ for every history.

\emph{Subtraction and alternative emission proposals.}
The optional correlated $n=0$ subtraction replaces $Z_e$ in
Eq.~\eqref{eq:qq-estimator} by $Z_e-Z_B$, with $Z_B=+4$ for the
octet and $-4$ for the singlet. The subtraction uses the same $t_s$
and no collinear Sudakov. It removes the zero-collinear-operator term,
rather than the subset of events with zero real emissions.
Figure~\ref{fig:comparison} shows the full tower.

\section{Monte Carlo algorithm for $qg\to qV$}
\label{sec:showeralgoqg}

\emph{Carried state.}
The insertion times and cumulative measurement are generated as
in the quark channel. During sampled evolution, one shared particle
list specifies the directions, and two connection lists $f_L,f_R$
specify the amplitude and conjugate amplitude. With $M$ connections
per flow, explicit inverse-colour power $p$ and signed weight $W$,
an unequal pair represents the correlated combination
\[
 W\Nc^{2-M-p}
 \bigl(|f_L\rangle\langle f_R|+\epsilon|f_R\rangle\langle f_L|\bigr),
 \qquad \epsilon=+1,\,-1,\,+1
 \quad\text{before, between and after the Glaubers}.
\]
For equal flows, the symmetric combination is included once and the
antisymmetric combination vanishes. One such combination is one
sampling alternative. The Born state and first real update are kept
as small exact sums, as specified below.

An endpoint is one fundamental or antifundamental colour
index of a particle. In the all-outgoing convention, incoming quark
$0$ has endpoint $0_{\bar3}$, outgoing quark $2$ has $2_3$, and each
gluon $i$ has both $i_3$ and $i_{\bar3}$. These labels specify colour
indices, independently of the particle directions $n_\pm$.

\emph{One complete event.}
\begin{enumerate}
\item Draw $t_1<t_2<t_s$ and evolve the Born density to $t_1$.
Keep the first real update as an exact sum; sample a colour alternative
at each subsequent emission. Histories with no real emission
before $t_1$ have zero weight.
\item If the next emission time $t_e$ lies below $t_2$, sample
the first Glauber at $t_1$ and continue real evolution up to $t_2$.
Otherwise, evaluate both Glauber colour actions in the final sum. Keep the already
drawn $t_e$: a Glauber does not change the scalar emission rate.
\item Evaluate the second Glauber's colour action exactly.
For an empty interval between the Glaubers, evaluate their ordered
product together. Then perform the soft contraction.
Combine this result with the time and operator
normalization factors below, and fill the cumulative bins with $T>t_s$.
\end{enumerate}

\emph{Born initialization.}
The colour-flow representation assigns a fundamental and an antifundamental
index to each gluon. Its physical adjoint component is obtained by
subtracting the trace,
$X_{ij}\mapsto X_{ij}-\delta_{ij}X_{kk}/\Nc$.
The disconnected trace in Eq.~\eqref{eq:qgborn} implements this subtraction
for the incoming gluon; the $L_g/\Nc$ term in $B_q$ implements it for
the emitted gluon. In these subtraction terms, the gluon's fundamental and
antifundamental ends are connected to each other, as in $(1,1)$ for
the incoming gluon. They are auxiliary colour tensors, which must be
combined coherently with the other flows to recover a traceless gluon.

\begin{figure}[!htbp]
 \centering
 \includegraphics[width=0.8\textwidth]{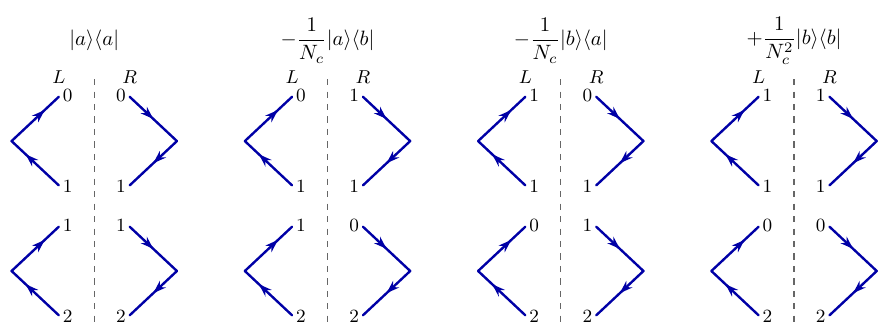}
 \caption{The four terms of the unnormalized Born density $|B\rangle\langle B|$.
The leading tensor $a$ connects $2\to1\to0$; in $b$, the two ends of
incoming gluon $1$ connect to each other and quark $2$ connects to $0$.
The normalization correction in Eq.~\eqref{eq:qg-born-normalized}
has the first diagram's connections and coefficient $1/\Nc^2$.
As in Fig.~\ref{fig:flow-history}, dashed lines separate amplitude and
conjugate amplitude, and arrows follow colour indices, reversing on
conjugation. They do not indicate particle directions.}
 \label{fig:qg-born}
\end{figure}

Let $|a\rangle$ and $|b\rangle$ be the leading chain and
subtraction tensor in Eq.~\eqref{eq:qgborn}.
Figure~\ref{fig:qg-born} shows their four density-matrix terms. The Born tensor has norm $\Nc^2-1$,
whereas our flow traces are divided by $\Nc^2$. We therefore multiply
its density by $\Nc^2/(\Nc^2-1)$ and expand to the retained order:
\begin{equation}
 \widetilde\rho_B=
 |a\rangle\langle a|
 -\frac{|a\rangle\langle b|+|b\rangle\langle a|}{\Nc}
 +\frac{|b\rangle\langle b|+|a\rangle\langle a|}{\Nc^2}
 +\text{terms beyond the retained colour order}.
 \label{eq:qg-born-normalized}
\end{equation}
The last $|a\rangle\langle a|/\Nc^2$ is the normalization
correction that makes the initial trace unity through relative
order $1/\Nc^2$.
The normalized Glauber operator
$G=V_G/(-4i\pi\Nc)$, defined in Eq.~\eqref{eq:qgglauber},
annihilates this coherent density: $G\widetilde\rho_B=0$.
Histories without a real emission before $t_1$ give zero.

\emph{Collinear emission times and colour updates.}
We use $\operatorname{tr}(t^at^b)=\delta^{ab}/2$ and convert
the emitted adjoint index to a colour-flow pair with $\sqrt2\,t^a_{ij}$.
The Fierz identity
\[
 \sum_a t^a_{ij}t^a_{kl}
 =\frac12\left(\delta_{il}\delta_{kj}
              -\frac1{\Nc}\delta_{ij}\delta_{kl}\right)
\]
then represents the charge insertion by $B_i/\sqrt2$, where
the endpoint maps $B_i$ are given in Eq.~\eqref{eq:realqg}.
Consequently, the real part of $-8\pi t\Gc$ acts as
\[
 32\pi t\sum_{i=0,1}
 (\bm T_{i,L}\circ\bm T_{i,R})\rho
 =16\pi t(B_q\rho B_q^\dagger+B_g\rho B_g^\dagger).
\]
The virtual coefficient is
$-32\pi t(C_F+C_A)=-48\pi\Nc t+16\pi t/\Nc$.
Dividing the real term by the leading scalar rate gives
the normalized map and the evolution generator:
\begin{equation}
 \mathcal R\rho=
 \frac{B_q\rho B_q^\dagger+B_g\rho B_g^\dagger}{3\Nc},\qquad
 -8\pi t\Gc=\lambda_{qg}(t)(\mathcal R-\mathbf1)
              +\frac{16\pi t}{\Nc}\mathbf1,\qquad
 \lambda_{qg}(t)=2\kappa_{qg}t=48\pi\Nc t.
 \label{eq:qg-generator}
\end{equation}
The factor $1/(3\Nc)=16\pi t/\lambda_{qg}(t)$ therefore
comes from dividing out the emission rate, whose leading colour
factor is $C_F+C_A\to3\Nc/2$. The leading scalar term gives
$\Delta_c(t_a,t_b)=\exp[-24\pi\Nc(t_b^2-t_a^2)]$.
Generate successive times by Eq.~\eqref{eq:sudakov}, using
$\kappa_{qg}$. The last term in Eq.~\eqref{eq:qg-generator}
is the subleading part of $C_F$. It does not change the two-Glauber
gap result at the retained colour order.

For a general input $|f_L\rangle\langle f_R|$, the two terms of
each $B_i$ give four products, hence the eight elementary operations
below. Each row replaces the flows by $A_Lf_L,A_Rf_R$, multiplies
the input coefficient by $a_{\rm elem}$ and changes $p$ to $p+\Delta p$.
These elementary factors are combined below to obtain the
sampling coefficients $a_j$.
The maps act on each side's existing connections. Both sides receive
the same new gluon, with direction $n_g$; the common $1/\Nc$ in
$\mathcal R$ is absorbed by $M\to M+1$.
\begin{center}
\begin{tabular}{ccccc}
\toprule
$n_g$ & $A_L$ & $A_R$ & $a_{\rm elem}$ & $\Delta p$\\
\midrule
$n_+$ & $I_{\bar0}$ & $I_{\bar0}$ & $+1/3$ & 0\\
$n_+$ & $I_{\bar0}$ & $L_g$ & $-1/3$ & 1\\
$n_+$ & $L_g$ & $I_{\bar0}$ & $-1/3$ & 1\\
$n_+$ & $L_g$ & $L_g$ & $+1/3$ & 2\\
\midrule
$n_-$ & $I_{1_3}$ & $I_{1_3}$ & $+1/3$ & 0\\
$n_-$ & $I_{1_3}$ & $I_{1_{\bar3}}$ & $-1/3$ & 0\\
$n_-$ & $I_{1_{\bar3}}$ & $I_{1_3}$ & $-1/3$ & 0\\
$n_-$ & $I_{1_{\bar3}}$ & $I_{1_{\bar3}}$ & $+1/3$ & 0\\
\bottomrule
\end{tabular}
\end{center}

To sample a real update, apply these operations to every term of
the carried combination, including its relative sign. Keep the
accumulated weight $W$ outside this local expansion. For each output,
count the lost loops $d$ and retain $h=d+p\leq2$, as in
Eq.~\eqref{eq:grade}. Add coefficients of identical outputs, which
must agree in particle directions, both connection lists and $p$.
Group the surviving terms with their transposes in the convention
above. The coefficient $a_j$ of each resulting combination
includes the elementary factors $a_{\rm elem}$ and the coefficients
of the input terms. For unequal flows, the two terms have
coefficients $a_j$ and $\epsilon a_j$,
with no extra factor of two. Explicit inverse-colour powers remain
in $p$ and are excluded from $a_j$.

Choose one alternative with $p_j=|a_j|/\sum_k|a_k|$, update
$W\to W a_j/p_j$, and adopt its directions, connections and $p$.
Thus the beam and colour connections are selected jointly. The
coefficients are recomputed at every update; if all vanish, the
history has zero weight.

Our implementation keeps the first-real state
$\mathcal R\widetilde\rho_B$ as an exact sum. Apply the next sampled
operator to this entire sum before selecting an alternative; if no
sampled update follows, contract it exactly. Like the exact terminal
sums, this numerical choice avoids additional colour sampling while
preserving the conditional mean.

For example, let the first emitted gluon be labelled $3$.
The two quark attachments on the leading Born component give
\[
 \begin{aligned}
 |u\rangle&=|\{(1,3),(2,1),(3,0)\}\rangle,\qquad
 |v\rangle=|\{(1,0),(2,1),(3,3)\}\rangle,\\
 B_q|a\rangle&=-|u\rangle+\frac1{\Nc}|v\rangle .
 \end{aligned}
\]
The first term inserts the gluon next to incoming quark $0$;
the second appends its disconnected trace. Their density is
\[
 B_q|a\rangle\langle a|B_q^\dagger
 =|u\rangle\langle u|
 -\frac{|u\rangle\langle v|+|v\rangle\langle u|}{\Nc}
 +\frac{|v\rangle\langle v|}{\Nc^2}.
\]
These are the four quark terms from this Born component, before
the common $1/(3\Nc)$ in $\mathcal R$. The initial update acts on
the complete Born density in Eq.~\eqref{eq:qg-born-normalized}.

\begin{figure}[!htbp]
 \centering
 \includegraphics[width=0.8\textwidth]{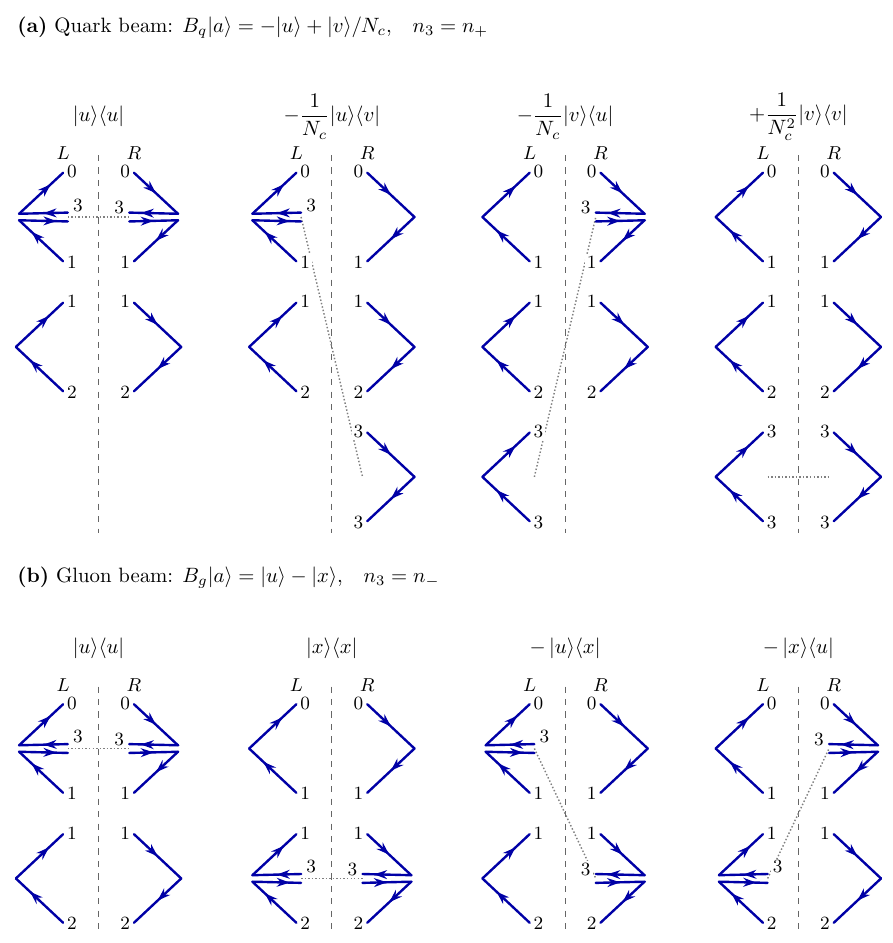}
 \caption{The eight elementary real-emission operations, illustrated
by their action on $|a\rangle\langle a|$: four from the quark beam (a)
and four from the gluon beam (b). These same eight operations apply
to every input flow pair, at every emission multiplicity, before
colour truncation and combination of identical outcomes.
The common factor $1/(3\Nc)$ is omitted. Particle $3$ is the emitted
gluon; dotted guides identify it on the two sides of the cut.
In $v$, its two colour ends are connected to each other, giving the
subtraction term in $B_q$. The negative terms in (b) attach the same
emitted particle to different ends of incoming gluon $1$ on the two
sides. For a general input $|f_L\rangle\langle f_R|$, the
attachments act on $f_L$ and $f_R$, respectively; their existing
connections determine the output flows.}
 \label{fig:qg-real}
\end{figure}

Figure~\ref{fig:qg-real} displays these four terms and the four
from the gluon beam. For the latter,
$B_g|a\rangle=|u\rangle-|x\rangle$, where
$|x\rangle=|\{(1,0),(2,3),(3,1)\}\rangle$.
The colour connections of $u$ are the same for either beam, but the
emitted particle has direction $n_+$ for the quark term and $n_-$
for the gluon term; these are distinct states.

\emph{Two Glauber insertions.}

Every history contains exactly two Glauber insertions, at $t_1$
and $t_2$, with the same operator $G$ in Eq.~\eqref{eq:qgglauber}.
This represents a virtual exchange: it acts on the existing
colour connections and their coefficients, without adding a gluon
to the particle list.
Its $P$ and $K$ terms reconnect incoming colour endpoints on either
side of the cut. If a real emission separates the Glaubers, select
one colour alternative for the first using the coefficients $a_j$
and the weight update in Eq.~\eqref{eq:signed}.
The second Glauber's colour terms are always summed exactly:
there is no selection with probability $p_j$ and no factor $a_j/p_j$
at this step. If no emission separates the Glaubers, sum both colour
actions exactly. The first Glauber always acts at $t_1$ and the second
at $t_2$; only the numerical evaluation of their colour terms differs.
Their insertion times are sampled in either case.
Exact summation preserves the expectation value and is convenient
when the resulting terms can be contracted directly with the soft
measurement.

For a sampled first Glauber, apply the four terms of
Eq.~\eqref{eq:qgglauber} to each component of the carried state,
using the same colour truncation, combination and signed selection
as above. For example, if one emission precedes the first Glauber
and another lies between the Glaubers, the first colour choice is
made from $G\mathcal R\widetilde\rho_B$.
Besides $u,v,x$ defined above, the resulting flows are
\[
 \begin{aligned}
 |y\rangle&=|\{(1,3),(2,0),(3,1)\}\rangle,\\
 |z\rangle&=|\{(1,1),(2,3),(3,0)\}\rangle.
 \end{aligned}
\]

\begin{figure}[!htbp]
 \centering
 \includegraphics[width=0.8\textwidth]{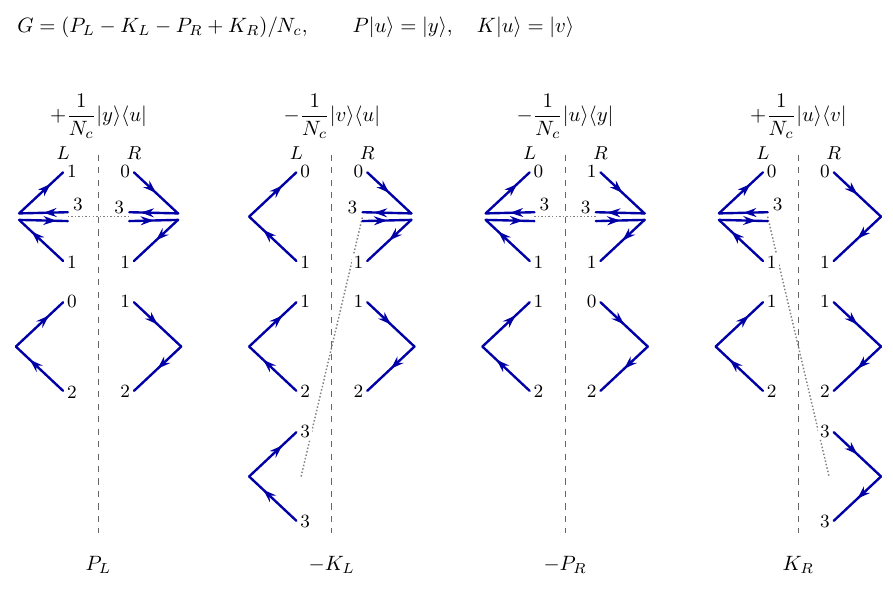}
 \caption{Action of $G$ on one diagonal pair $|u\rangle\langle u|$.
The first two terms change the left flow; the last two change the right.
The $P$ terms exchange incoming anticolour labels $0,1$.
The $K$ terms replace $(1,3),(3,0)$ by $(1,0),(3,3)$.
The common physical phase $-4i\pi\Nc$ is not included.
The four terms form two antisymmetric transpose combinations.}
 \label{fig:qg-glauber-action}
\end{figure}

To see how these arise, start with $|u\rangle\langle u|$.
Swapping the incoming anticolour ends gives $P|u\rangle=|y\rangle$;
joining ends $1_3$ and $0_{\bar3}$ gives $K|u\rangle=|v\rangle$.
Thus the four terms of $G$ give
\[
 G\bigl(|u\rangle\langle u|\bigr)
 =\frac{|y\rangle\langle u|-|v\rangle\langle u|
        -|u\rangle\langle y|+|u\rangle\langle v|}{\Nc}.
\]
Figure~\ref{fig:qg-glauber-action} shows this update.
Acting on $x$ instead gives $P|x\rangle=|z\rangle$ and
$K|x\rangle=\Nc|x\rangle$, since $(1,0)$ is already connected.
Apply the same four operations to every term of the exact first-real
state before combining and selecting alternatives.

\begin{figure}[!htbp]
 \centering
 \includegraphics[width=0.8\textwidth]{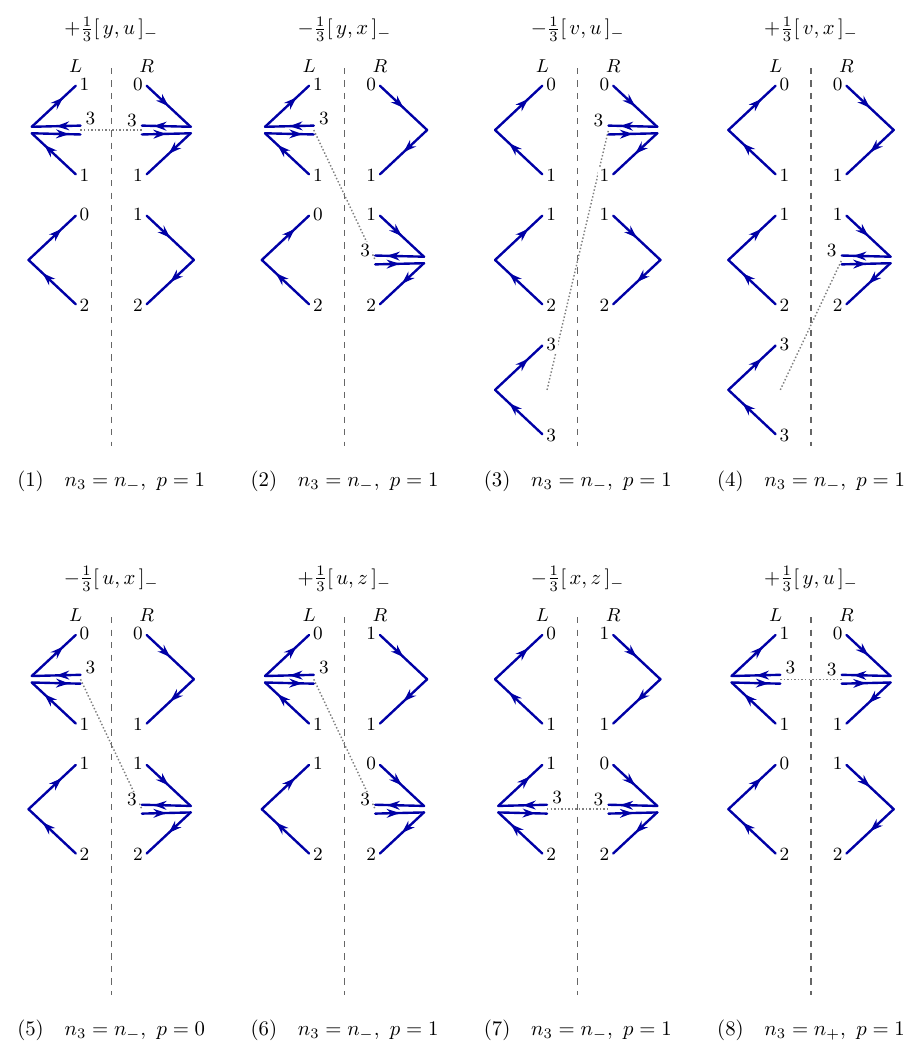}
 \caption{The eight alternatives of the sampled first Glauber update
$G\mathcal R\widetilde\rho_B$, in the same order as the table.
Here $[f,g]_-\equiv|f\rangle\langle g|-|g\rangle\langle f|$;
each panel draws its first pair, and its exchanged pair is retained
with the opposite sign. Coefficients are $a_j$; the explicit power
$\Nc^{-p}$ is tracked separately. Rows 1 and 8 have identical colour
connections but different emitted-gluon directions. All alternatives
have $d+p=2$ and probability $1/8$.}
 \label{fig:qg-glauber-alternatives}
\end{figure}

After applying $G$ to the complete first-real state, truncating
and combining identical outcomes, this example has the eight
retained alternatives listed below. Each row represents
$|f_L\rangle\langle f_R|-|f_R\rangle\langle f_L|$ with coefficient
$a_j$ and explicit power $p$. All have $h=2$; $n_g$ is the direction
of emitted gluon $3$.
\begin{center}
\begin{tabular}{ccccc}
\toprule
$n_g$ & $f_L$ & $f_R$ & $p$ & $a_j$\\
\midrule
$n_-$ & $y$ & $u$ & 1 & $+1/3$\\
$n_-$ & $y$ & $x$ & 1 & $-1/3$\\
$n_-$ & $v$ & $u$ & 1 & $-1/3$\\
$n_-$ & $v$ & $x$ & 1 & $+1/3$\\
$n_-$ & $u$ & $x$ & 0 & $-1/3$\\
$n_-$ & $u$ & $z$ & 1 & $+1/3$\\
$n_-$ & $x$ & $z$ & 1 & $-1/3$\\
$n_+$ & $y$ & $u$ & 1 & $+1/3$\\
\bottomrule
\end{tabular}
\end{center}
Figure~\ref{fig:qg-glauber-alternatives} depicts these eight
combined alternatives; their number is specific to this input and
operator. Thus $\sum_j|a_j|=8/3$ and each alternative has probability
$1/8$. Selecting, for example, the third row multiplies the running
weight by $-8/3$ and retains $p=1$ together with
$|v\rangle\langle u|-|u\rangle\langle v|$ and $n_g=n_-$.
Colour truncation removes terms beyond the retained accuracy;
combining identical outputs can also produce exact cancellations.
For example, $K|x\rangle=\Nc|x\rangle$, so the $-K_L$ and $+K_R$
terms in $G(|x\rangle\langle x|)$ cancel.

\emph{Final measurement and event weight.}
At this stage, evolution ends. Keep the accumulated weight $W$
fixed. If the first Glauber's colour action has already been applied
through sampling, expand the second Glauber; otherwise expand both
Glaubers in order. Acting on
the carried colour combination, including its transpose (or the exact
first-real sum), this produces a signed sum of temporary pairs $(F_L,F_R)$,
each with its coefficient and power $p$. Their connections can differ
from the carried $(f_L,f_R)$. Apply the soft operator
and take the trace for each pair, then add the resulting numbers.
Their sum multiplies $W$ to give the final colour coefficient.
No output pair is selected or propagated further.

After both Glaubers, the soft operator is
$S=\mathcal S_{\rm gap}/(\Nc\Delta Y)
=8\sum_{i<j,n_i\ne n_j}\bm T_i\cdot\bm T_j/\Nc$.
The angular integral is included in $\Delta Y$.

For each temporary pair $(F_L,F_R)$, sum over all particle pairs
with opposite directions and all pairs of their endpoints. Each soft
term changes $F_L$ to $F'_L$, keeping that pair's $F_R$ unchanged,
according to
\begin{equation}
 \bm T_e\cdot\bm T_f=
 \begin{cases}
 \tfrac12(P_{ef}-\mathbf1/\Nc),&\text{ends of the same type},\\
 -\tfrac12(K_{ef}-\mathbf1/\Nc),&\text{ends of opposite types}.
 \end{cases}
 \label{eq:fierz}
\end{equation}
Here $P_{ef}$ exchanges partners; $K_{ef}$ joins the two ends
and reconnects their former partners. If they are already connected,
$K_{ef}$ leaves the flow unchanged and supplies $\Nc$.
The identity operator $\mathbf1$ leaves all connections unchanged.
The endpoint types determine whether $P$ or $K$ occurs in a given
term.
These are the same elementary reconnections that enter $G$,
now used to evaluate the soft operator after both Glaubers.
Here they act on all opposite-direction particle pairs, and every
contribution is contracted and summed without sampling.

\begin{figure}[!htbp]
 \centering
 \includegraphics[width=0.8\textwidth]{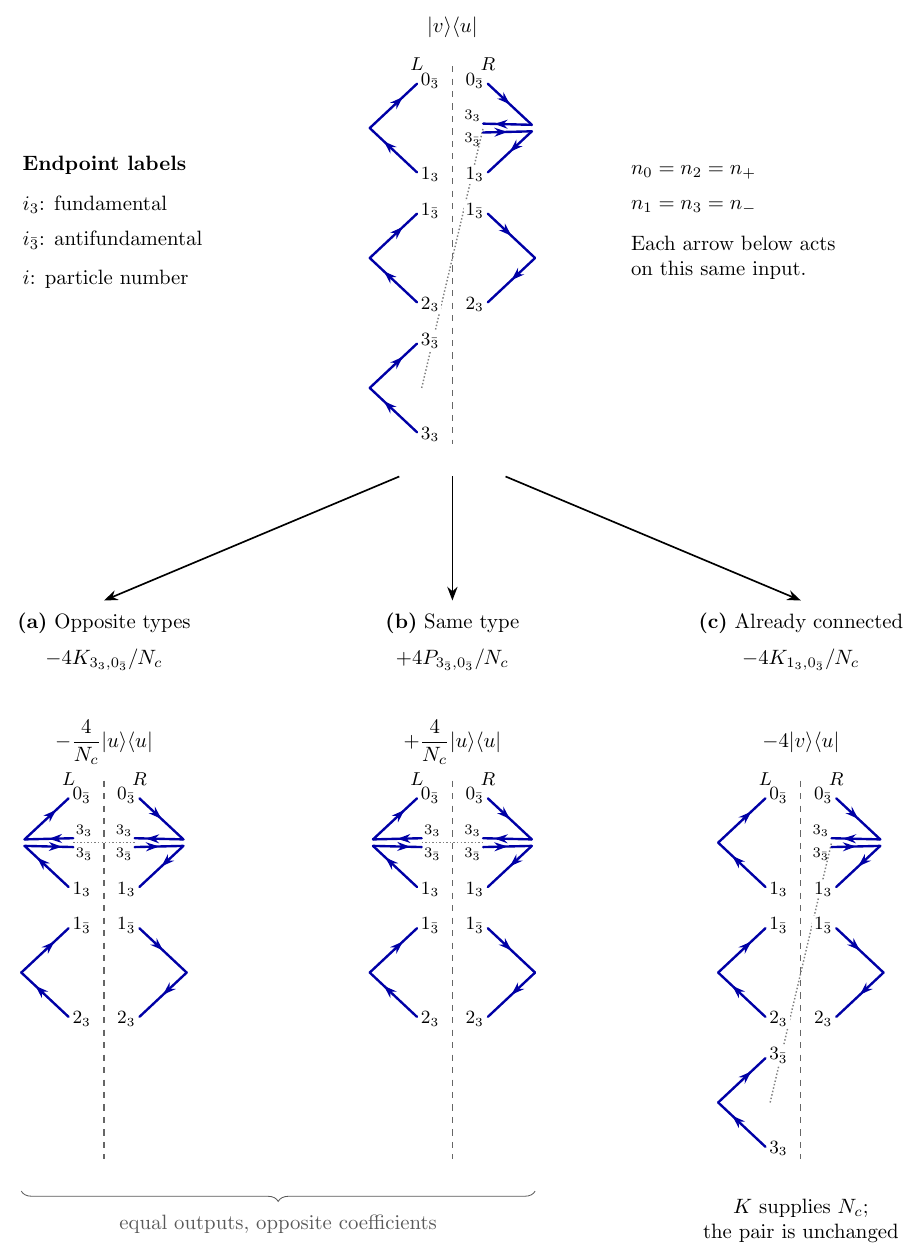}
 \caption{Soft action on $|v\rangle\langle u|$, with $n_3=n_-$,
showing both amplitude ($L$) and conjugate amplitude ($R$).
(a) The opposite-type pair $(3_3,0_{\bar3})$ gives
$-4K_{3_3,0_{\bar3}}/\Nc$; (b) the same-type pair
$(3_{\bar3},0_{\bar3})$ gives $+4P_{3_{\bar3},0_{\bar3}}/\Nc$.
Both turn $v$ into $u$, with opposite coefficients.
(c) Ends $1_3,0_{\bar3}$ are already connected: $K$ supplies $\Nc$,
giving $-4|v\rangle\langle u|$.
Identity terms are not drawn: they cancel between (a,b) and are
beyond the retained order for this input in (c).
The arrows show separate contributions to an explicit sum;
no branch is sampled.}
 \label{fig:qg-endpoints}
\end{figure}

For a separate terminal example, take one emission before $t_1$
and none between the Glaubers. Both Glaubers are then summed
exactly. The resulting $G^2\mathcal R\widetilde\rho_B$ contains
$|v\rangle\langle u|$ with $p=1$, $n_3=n_-$ and coefficient $+1/3$.
For this input, the two endpoint pairs of particles $(3,0)$ give
\[
 \frac8{\Nc}\bm T_3\cdot\bm T_0
 =\frac4{\Nc}\bigl(P_{3_{\bar3},0_{\bar3}}
                  -K_{3_3,0_{\bar3}}\bigr),
\]
where their identity terms have cancelled.
Figure~\ref{fig:qg-endpoints} shows these two contributions and an
already-connected example. All are added with their signs.

Next, contract each output $|F'_L\rangle\langle F_R|$ by
identifying matching external indices across the cut. Each closed
index loop supplies $\Nc$. If the soft term adds an explicit power
$r$, let $c$ include its numerical coefficient and that of the input
term in the terminal sum. Its contribution is
\[
 \frac{W\Nc^{2-M-p}}{\Nc^2}\,
       \frac{c}{\Nc^r}\,\Nc^{M-d'}
       =Wc\,\Nc^{-(p+r+d')},\qquad
 d'=M-\operatorname{cycles}(F_R^{-1}F'_L).
\]
Summing these scalar contributions gives $h_0+h_2/\Nc^2$.
For the complete two-Glauber gap trace, $h_0=0$, and
\[
 h_2=W\!\sum_{\substack{\text{terminal terms}\\p+r+d'=2}}c.
\]
This sum includes the correlated transpose, all remaining
Glauber terms and all soft endpoint contributions, with their signs.
Thus $h_2$ already contains $W$; the event normalization below must
not multiply by $W$ again. Both equal and unequal flows contribute:
the condition $p+r+d'=2$ allows
$(d',p+r)=(0,2),(1,1),(2,0)$.

\begin{figure}[!htbp]
 \centering
 \includegraphics[width=0.85\textwidth]{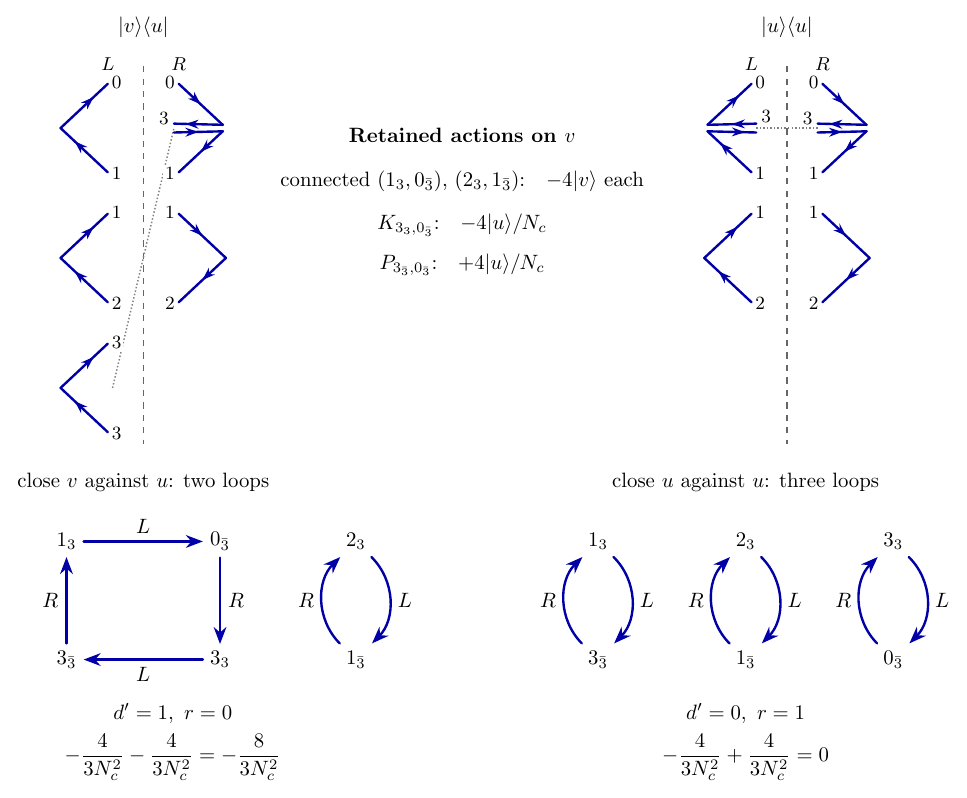}
 \caption{Final trace for the example in Fig.~\ref{fig:qg-endpoints},
with input coefficient $+1/3$ and explicit power $p=1$.
The bottom diagrams follow alternating left and right connections
after identifying matching indices across the cut; each closed path
gives $\Nc$. The two connected endpoint terms leave $v$ unchanged,
giving two loops ($d'=1$) and a total $-8/(3\Nc^2)$.
The two reconnections give the diagonal pair $|u\rangle\langle u|$
and three loops ($d'=0$), but cancel in the signed sum.
The transpose of the input pair gives the same total.}
 \label{fig:qg-soft-contraction}
\end{figure}

Figure~\ref{fig:qg-soft-contraction} completes the terminal
example above. The connected endpoint pairs $(1_3,0_{\bar3})$ and $(2_3,1_{\bar3})$
each give $-4|v\rangle$. Closing against $u$ gives two loops,
$d'=1$, hence $-4/(3\Nc^2)$ per term.
The two reconnections in Fig.~\ref{fig:qg-endpoints}(a,b) instead
give $|u\rangle\langle u|$, with three loops and coefficients
$\mp4/\Nc$; they cancel. The input pair therefore contributes
$h_2=-8/3$, and its transpose contributes the same amount.

\FloatBarrier
Finally, restore the physical operator and time normalization.
The two Glauber factors multiply to
$(-4i\pi\Nc)^2=-16\pi^2\Nc^2$: thus the minus sign.
The soft operator supplies $\Nc\Delta Y$ and the retained trace
$h_2/\Nc^2$. All signs from colour reconnections and the gap
measurement are already included in $h_2$.
We find

\begin{equation}
 X_e(T)=-\frac{T_{\max}^3}{6}\,16\pi^2\Nc\Delta Y\,
                       h_{2,e}\,\Theta(T-t_s).
 \label{eq:qg-estimator}
\end{equation}
Its mean estimates $C_{qg}(T)$ as defined in
Sec.~\ref{app:BNSSqg}. No $n=0$ subtraction is needed.
A low-order check gives $h_2=-64/3$ for one normalized real map,
two $G$ maps and $S$. Including the first-emission density
$\lambda_{qg}(t)$ and ordered insertion integrals reproduces
Eq.~\eqref{eq:qg-smallw}.

\FloatBarrier
\emph{Sudakov accuracy.}
The Sudakov used to generate emissions already gives the complete
virtual contribution needed for our retained two-Glauber gap result.
The subleading part of $C_F$ supplies a scalar correction of relative
order $1/\Nc^2$ at fixed $\Nc t^2$. Since the gap trace starts at
$1/\Nc^2$, this factor first changes it at $1/\Nc^4$.
No additional Sudakov reweighting is needed for the results in
Fig.~\ref{fig:comparison}.

\section{Numerical comparison}\label{sec:numerics}

Figure~\ref{fig:comparison} uses ten independent runs for each channel,
with $10^7$ events in total for each $qq'$ colour state and $5\times10^7$
for $qg$. The full gap width is $\Delta Y=2$. At $\xi=8$, Marzili gives
\begin{equation}
 \begin{aligned}
 C_O&=6.2759(47),& C_S&=-11.6853(45),& C_{qg}&=0.1099(247),
 \end{aligned}
 \label{eq:endpoint}
\end{equation}
where parentheses denote one standard error on the final digits.
The corresponding analytic values are $6.2754$, $-11.6804$ and $0.0971$.
Errors are obtained from the sample variance of the event
contributions, including zero-weight events, combined over the
independent runs.
The cumulative points share histories, so their deviations from the
reference curves are correlated.
At this endpoint, negative contributions account for $15.5\%$ of
nonzero octet events and $8.2\%$ of nonzero $qg$ events. The weight magnitudes
also matter: for $qg$, the mean absolute contribution is about $52$ times
the magnitude of the mean signed contribution. This explains the strong
cancellations despite the smaller negative-event fraction. All nonzero
singlet contributions are negative, reflecting the sign of this correction;
there is no cancellation between opposite signs within that sample.

\end{document}